# Optical dipole force on ladder-like three-level atomic systems by chirped few-cycle-pulse laser fields

**Parvendra Kumar and Amarendra K. Sarma***
Department of Physics, Indian Institute of Technology Guwahati, Guwahati-781039, Assam, India.
*Electronic address: aksarma@iitg.ernet.in

We report a study on the optical dipole force in a ladder like three level atomic systems in the context of coherent population transfer with the chirped few-cycle-pulse laser fields. The phenomenon of coherent population transfer is investigated by numerically solving the appropriate density matrix equations beyond the rotating wave approximation. On the other hand, optical dipole force is calculated by numerically solving the force equation and density matrix equations self-consistently. By analyzing the centre-of-mass motion, it is shown that the optical dipole force with chirped pulses may be used for focusing and defocusing of atoms in an atomic beam similar to the near or non-resonant optical dipole force. Moreover, the robustness of the population transfer against the variation of the pulse parameters and the effect of the variation of the Rabi frequencies and the chirp rates on the optical dipole force are also investigated. The proposed scheme may open new perspectives in the focusing and de-focusing of atoms and molecules since this scheme mitigates the demand for the generation of transform-limited pulse laser fields at arbitrary frequencies.

## I. Introduction

Manipulation and coherent control of atoms and molecules, and various schemes for laser cooling and trapping by using optical force is of tremendous importance to many fields of physics and chemistry [1-6]. Recently, there has been a resurgence of interest in the so-called light or optical force due to the recent progress in the generation of well controlled femtosecond laser pulses [7-10]. Generally, there are two kinds of optical force: the reactive or the so-called optical dipole force and the dissipative or the so-called spontaneous force [11]. The optical dipole force arises from the interaction between the induced dipole moment and the gradient of the electric field envelope while the dissipative force arises from the impulse experienced by an atom when it absorbs or emits a quantum of photon momentum [12]. Several authors have utilized optical force, through the linearly chirped laser pulses, for slowing down, acceleration [13] and forced rotation [14] of molecules. It is worthwhile to mention that in late seventies, in an experimental study, Bjorkholm et al. [15] demonstrated the phenomena of focusing, defocusing and steering of neutral sodium atoms by using CW laser fields and showed that atoms could be expelled from the laser beam due to the so called optical dipole force. Recently, the phenomena of focusing, defocusing, and steering of the neutral atoms in the few-cycle-pulse laser field are theoretically analyzed by the authors [16]. The creation of optical lens for atomic and molecular beam by optical dipole force has also been demonstrated by several authors [17-19]. It may be noted that, the spontaneous force can be used for cooling the atoms but cannot be used for trapping the cold atoms. This may be attributed to the fact that the upper limit of the dissipative forces is limited by saturation effects due to the spontaneous emission. Recently, observation of very strong optical force, so-called stimulated force produced by coherent exchange of momentum between atoms and light fields implemented by adiabatic rapid passage (ARP) and bi-chromatic pulses is reported [20- 25]. Here, it is worth to mention that the so-called optical dipole force with ARP scheme is not explored by any authors to the best of our knowledge. In the ARP scheme, the laser pulse frequency is swept through a static resonance of atom or molecule. It is worthwhile to mention that ARP and the so-called stimulated Raman adiabatic passage (STIRAP) are the most widely used schemes for controlling the population transfer between the quantum states of atoms and molecules. These have found many potential applications in spectroscopy, collision-dynamics and control of chemical reactions etc. [26-29]. The effect of Doppler broadening of transition lines in STIRAP could be minimized by increasing the intensity of the pulses [30]. However due to increase in intensity, the Rabi frequency may exceed the adjacent transition frequency leading to the poor selectivity of the population transfer with STIRAP scheme [31].The optical dipole force on the atoms with STIRAP scheme may not be as effective as the one with the chirped few-cycle-pulse laser fields, owing to the resonant interaction in case of STIRAP. In this work we report a study on the optical dipole force, in the ladder like three level atomic systems in the context of coherent population transfer, with a scheme similar to the so-called adiabatic rapid passage (ARP) scheme. We show a nearly complete population transfer in atoms, co-propagating with linearly chirped few-cycle-pulse laser fields through judicious choice of the laser pulse parameters. For the chosen pulse parameters, we calculate the optical dipole force on moving atoms by numerically solving the density matrix equations and force equation self-consistently. In Sec. II we present the density matrix equations that describe the interaction of the ladder-like three level atomic systems with the linearly chirped few-cycle laser pulses. Force equation is also presented. Sec. III contains our simulated results and discussions followed by conclusions in Sec. IV.

## II. THE MODEL

The sketch of our scheme for the calculation of optical dipole force on sodium atoms is depicted in Fig. 1. We consider a ladder-like atomic system interacting with two few-cycle-pulse laser fields. In this work the states $|1\rangle$, $|2\rangle$ and $|3\rangle$ refers to the $3s_{1/2}, 3p_{3/2}$ and $4s_{1/2}$ quantum states of neutral sodium atoms.

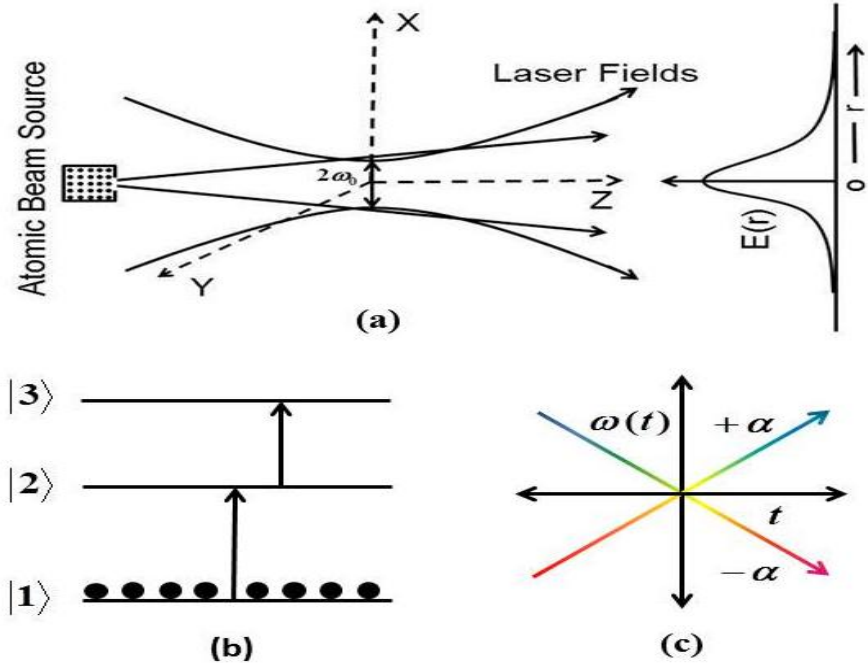

Fig. 1(a) Sketch of the proposed scheme for the calculation of optical dipole force on an atomic beam co-propagating with pulsed laser fields (b) Three level atomic system (c) Time dependent frequency of up-chirped ($+\alpha$) and down-chirped ($-\alpha$) pulses.

The electric field of the collimated linearly polarized laser interacting between $|1\rangle$ and $|2\rangle$ is given by $\vec{E}_1 = \vec{A}_1(r,t)\cos(\omega_{p1}t - \phi_1(z))$, where $\vec{A}_1(r,t)$, and $\omega_{p1} = \omega_{10} + \alpha_1 t - \omega_{D1}$ are, respectively, the field envelope and the time dependent frequency of the pulse. $\omega_{10}$ is the central frequency of the pulse, $\omega_{D1} = \vec{k}_1 \cdot \vec{v}$ refers to the detuning due to the Doppler shift of the transition lines of the atom moving with velocity $\vec{v}$ while $\alpha_1$ is the chirp rate. Exactly analogous expression for the linearly polarized laser pulse interacting between $|2\rangle$ and $|3\rangle$ is given by $\vec{E}_2 = \vec{A}_2(\vec{r},t)\cos(\omega_{p2}t - \phi_2(z))$, with $\omega_{p2} = \omega_{20} + \alpha_2 t - \omega_{D2}$ and $\omega_{D2} = \vec{k}_2 \cdot \vec{v}$. Here $\vec{k}_1$ and $\vec{k}_2$ are the wave vectors of pulsed laser fields. On the other hand, $\phi_1(z) = k_1 z$ and $\phi_2(z) = k_2 z$ are the corresponding longitudinal phases. The respective time dependent detunings from the atomic resonances are defined as: $\Delta_1(t) = \omega_{21} - \omega_{p1}$ and $\Delta_2(t) = \omega_{32} - \omega_{p2}$. For the Gaussian-shaped few-cycle laser fields, $\vec{A}_1(r,t) = \hat{\eta} A_{10} \exp\left[-\{(r/\omega_0)^2 + (t/\tau)^2\}\right]$ and $\vec{A}_2(r,t) = \hat{\eta} A_{20} \exp\left[-\{(r/\omega_0)^2 + (t/\tau)^2\}\right]$. Here, $\omega_0$ and $\tau_p = 1.177\tau$ are the beam waist and pulse duration respectively. The density matrix equations, without invoking the so called rotating wave approximation, describing the temporal evolution of the density matrix elements, are:

$$\frac{d\rho_{11}}{dt} = i(\Omega_{12}\rho_{21} - \Omega_{21}\rho_{12})$$

$$\frac{d\rho_{22}}{dt} = i(\Omega_{12}(\rho_{12} - \rho_{21}) + \Omega_{23}(\rho_{32} - \rho_{23}))$$

$$\frac{d\rho_{33}}{dt} = i(\Omega_{32}\rho_{23} - \Omega_{23}\rho_{32}) \quad (1)$$

$$\frac{d\rho_{21}}{dt} = -i\omega_{21}\rho_{21} + i(\Omega_{12}(\rho_{11} - \rho_{22}) + \Omega_{23}\rho_{31})$$

$$\frac{d\rho_{32}}{dt} = -i\omega_{32}\rho_{32} + i(\Omega_{32}(\rho_{22} - \rho_{33}) - \Omega_{12}\rho_{31})$$

$$\frac{d\rho_{31}}{dt} = -i\omega_{31}\rho_{31} + i(\Omega_{23}\rho_{21} - \Omega_{12}\rho_{32})$$

Here, $\Omega_{12} = \Omega_{21} = \mu_{12} E_1(r,t)/\hbar$ and $\Omega_{23} = \Omega_{32} = \mu_{23} E_2(r,t)/\hbar$ are the time dependent Rabi frequencies for the transition with electric dipole moment $\mu_{12}$ and $\mu_{23}$, respectively. It should be noted that $\omega_{ij} = \omega_i - \omega_j$ and $\rho_{ij} = \rho_{ji}^*$. Using an approach based on the density matrix equations and the Ehrenfest's theorem, we derived the following expression for the optical dipole force [16, 31]:

$$\vec{F}_r = \mu_{12} u\{[\vec{\nabla} A_1(\vec{r},t)]\cos(\omega_{p1}(t)t - \phi_1(z))\} + \mu_{23} v\{[\vec{\nabla} A_2(\vec{r},t)]\cos(\omega_{p2}(t)t - \phi_2(z))\} \quad (2)$$

Here $u = (\rho_{21} + \rho_{12})$ and $v = (\rho_{32} + \rho_{23})$.

## III. RESULTS AND DISCUSSIONS

We solve Eq. (1) and Eq. (2) numerically using a standard fourth-order Runge-Kutta method. We assume that initially all the atoms are in the ground state $|1\rangle$. We use the following typical parameters for simulation: $\omega_{21} = \omega_{10} = 3.19\,\text{rad/fs}$, $\omega_{32} = \omega_{20} = 1.65\,\text{rad/fs}$, $\Omega_{21} = \Omega_{32} = 1.30$ rad/fs, $\alpha_1 = \alpha_2 = \pm 0.02\,\text{fs}^{-2}$, $\mu_{12} = \mu_{23} = 1.85\,Cm$ [32], $r = 70.71\,\mu m$, $\omega_0 = 100\,\mu m$, $v_z = 100\,m/s$ in the direction of pulse laser fields. The temporal pulse width is taken to be $\tau_p = 23.5\,fs$.

From Fig. 1(c) it could be seen that for the up-chirped pulses $\Delta_1(t) = \Delta_2(t) = +ve$ for the first half and $\Delta_1(t) = \Delta_2(t) = -ve$ for the second half of the pulses and vice-versa for the down-chirped pulses. In the present study, the centre of mass is taken as, $M = nm$. Here, $n = 10$, is the number of sodium atoms and $m = 22.98\,amu$, is the mass of one sodium atom. Temporal evolution of the populations $\rho_{11}$, $\rho_{22}$ and $\rho_{33}$ is depicted in Fig. 2.

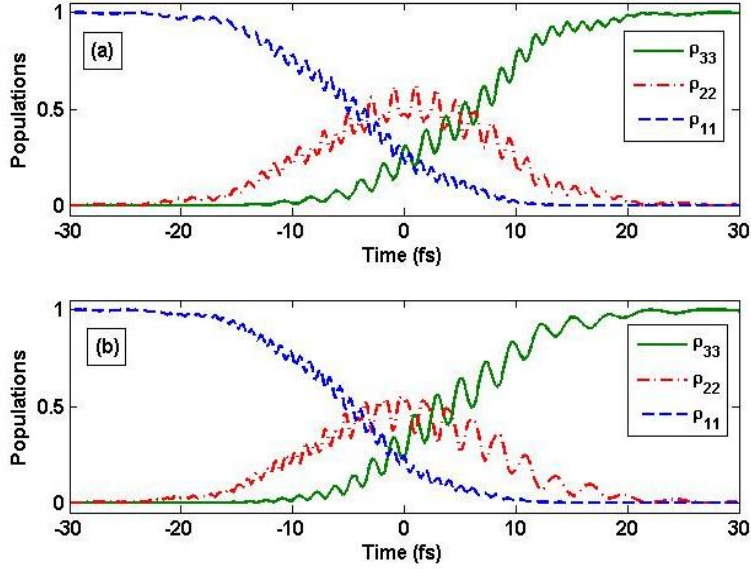

Fig. 2 (a) Temporal evolution of populations with the up-chirped few-cycle-pulse laser fields (b) Temporal evolution of populations with the down-chirped few-cycle-pulse laser fields.

It is clear from Fig. 2 that one can obtain almost complete population transfer (99.54 % with up-chirped and 99.45 % with down-chirped pulses) from the ground state $|1\rangle$ to the state $|3\rangle$. The ARP scheme is known to be robust against the variation of the laser pulse parameters in the case of many cycle pulses. It may be interesting to check the robustness of the scheme for the case of few-cycle pulses. Fig. 3 depicts the final population of the quantum state $|3\rangle$ against the variation of the Rabi frequencies, the spatial location of atoms in ensemble, the temporal pulse width and the chirp rate.

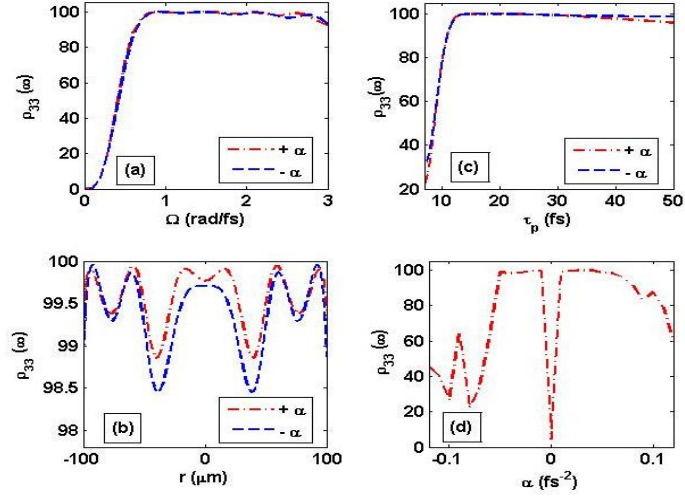

Fig. 3 Final population of quantum state $|3\rangle$ with up-chirped and down-chirped pulses, as a function of (a) Rabi frequencies, $\Omega_{12} = \Omega_{23} = \Omega$ (b) Temporal pulse width $\tau_p$ (c) Spatial location (r) of atoms in the ensemble (d) Chirp rates $\alpha_1 = \alpha_2 = \alpha$.

It is clear from the Fig. 3 that the final population $\rho_{33}(\infty)$ is highly robust against the variations of pulse parameters. In particular, it could be observed from the Fig. 3(b) that $\rho_{33}(\infty)$ is extremely robust against the variation of spatial location of atoms, i.e. all the atoms in ensemble are in the quantum state $|3\rangle$. In addition, it may be understood intuitively from Fig. 5(b) that the coherent population transfer with few-cycle-pulse laser fields is extremely insensitive to the Doppler broadening of the transition lines. Next, we study the optical dipole force on atoms during the coherent population transfer for the same set of simulation parameters as that in figure 2. In Fig. 4, we depict the temporal evolution of optical dipole force for both up-chirped and down-chirped pulses.

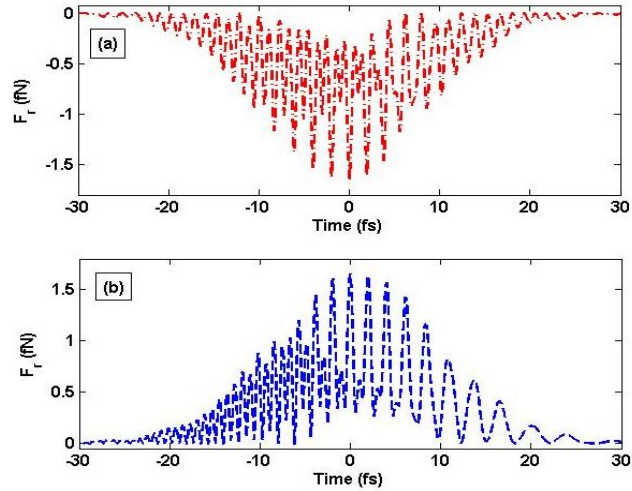

Fig. 4 Temporal evolution of optical dipole force with (a) up-chirped ($\alpha_1 = \alpha_2 = \alpha = 0.02\ fs^{-2}$) pulses (b) down-chirped ($\alpha_1 = \alpha_2 = \alpha = -0.02\ fs^{-2}$) pulses.

We find that the direction of the optical dipole force does not depend on the time dependent detuning from the atomic resonances and depends only on the direction of the chirp for the chosen parameters, as could be observed from Fig. 4(a). The force is negative with up-chirped pulses which in turn may lead to the focusing of atoms, co-propagating with laser pulses. On the other hand, as could be observed from Fig. 4(b), the optical dipole force with down-chirped pulses is positive, leading to the defocusing of the atoms. It is worth noting that for our chosen simulation parameters, the magnitude of the optical dipole force is much larger than that of the one considered by Bjorkholm et al. In Fig. 5, we depict the radial variation of optical dipole force and the effect of the Doppler broadening on the magnitude of optical dipole force.

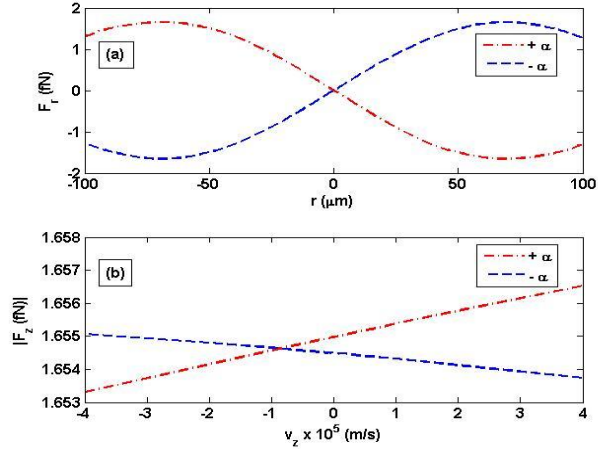

Fig. 5 Evolution of optical dipole force as a function of the (a) radial position of the atoms with up-chirped ( $\alpha_1 = \alpha_2 = \alpha = 0.02\,fs^{-2}$ ) and down-chirped ( $\alpha_1 = \alpha_2 = \alpha = -0.02\,fs^{-2}$ ) pulses. (b) Atomic velocity, $v_z$ in the direction of laser pulses with up-chirped ( $\alpha_1 = \alpha_2 = \alpha = 0.02\,fs^{-2}$ ) and down-chirped ( $\alpha_1 = \alpha_2 = \alpha = -0.02\,fs^{-2}$ ) pulses.

We observe, from Fig. 5(a), that the magnitude of optical dipole force is different at different radial position. Hence the atoms at different radial position would experience different magnitude of the optical dipole force for both up-chirped and down-chirped pulses. Also, it may be noted, from Fig. 5(b), that the magnitude of the optical dipole force is almost independent of the velocity of the atoms. In Fig. 6 the temporal evolution of the optical dipole force at two different values of the Rabi frequencies, for both up-chirped and down-chirped pulses, is shown.

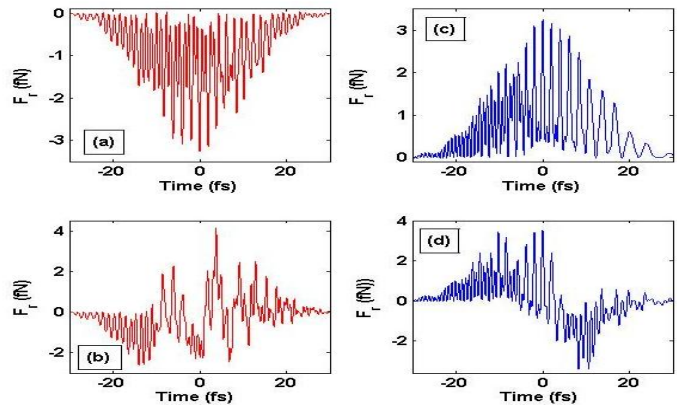

Fig. 6 Temporal evolution of optical dipole force as a function of Rabi frequencies (a) $\Omega_{12} = \Omega_{23} = 2.60\,rad/fs$ and $\alpha_1 = \alpha_2 = 0.02\,fs^{-2}$. (b) $\Omega_{12} = \Omega_{23} = 3.90\,rad/fs$ and $\alpha_1 = \alpha_2 = 0.02\,fs^{-2}$. (c) $\Omega_{12} = \Omega_{23} = 2.60\,rad/fs$ and $\alpha_1 = \alpha_2 = -0.02\,fs^{-2}$. (d) $\Omega_{12} = \Omega_{23} = 3.90\,rad/fs$ and $\alpha_1 = \alpha_2 = -0.02\,fs^{-2}$. Here all other simulation parameters are constant.

It could be observed that the magnitude of the optical dipole force increases with increase in the Rabi frequencies. However, for higher Rabi frequency $\Omega_{12} = \Omega_{23} > \omega_{21}$ the optical dipole force becomes oscillatory, as could be seen from Fig. 6(b) and 6(d), similar to that of the far-off resonant optical dipole force [16]. We find that, from Fig. 6(a) and 6(c), for Rabi frequencies $\Omega_{12} = \Omega_{23} < \omega_{21}$, the direction of the optical force depends on the direction of the chirp rates. The effect of chirp rate on the temporal evolution of the optical dipole force is depicted in Fig. 7.

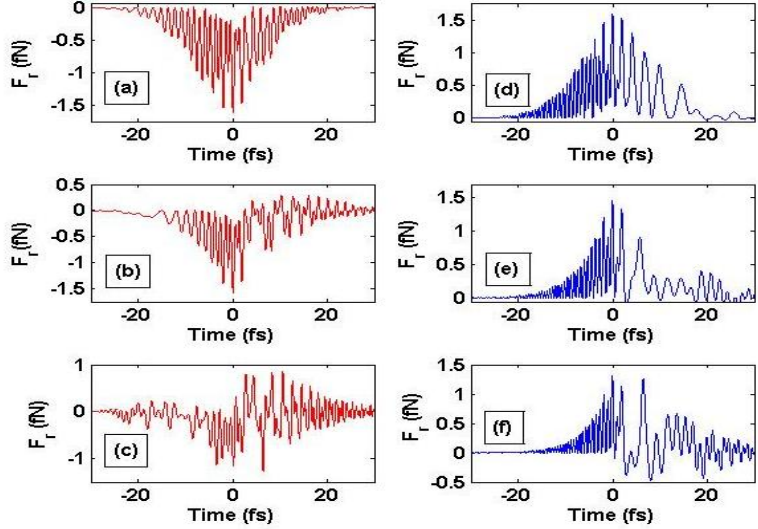

Fig. 7 Temporal evolution of optical dipole force as a function of Chirp rates (a) $\alpha_1 = \alpha_2 = 0.04\, fs^{-2}$ (b) $\alpha_1 = \alpha_2 = 0.08\, fs^{-2}$ (c) $\alpha_1 = \alpha_2 = 0.12\, fs^{-2}$ (d) $\alpha_1 = \alpha_2 = 0.04\, fs^{-2}$ (e) $\alpha_1 = \alpha_2 = -0.08\, fs^{-2}$ (f) $\alpha_1 = \alpha_2 = -0.12\, fs^{-2}$

It could be observed from Fig. 7 that the temporal evolution of the optical dipole force is sensitive to the chirp rates. It is clear from Fig. 7(c) and 7(f) that the optical dipole force becomes highly oscillatory for larger chirp rates and may not be relevant for practical applications. Finally, in Fig. 8 below, we show the trajectory of atoms subjected to the optical dipole force with Rabi frequencies, $\Omega_{12} = \Omega_{23} = 2.60\, rad/fs,$ rest of the parameters being the same as that of Fig. 4.

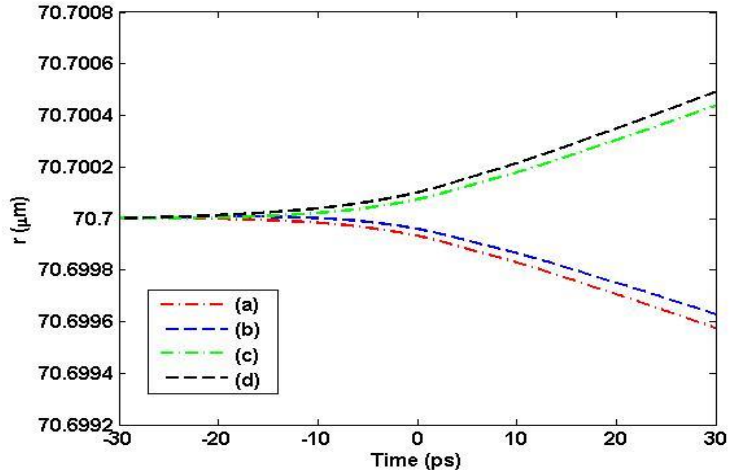

Fig. 8 Atomic motion under the influence of optical dipole force (a) $v_r = 10\,cm/s, \alpha = 0.02\,fs^{-2}$ (b) $v_r = 100\,cm/s, \alpha = 0.02\,fs^{-2}$ (c) $v_r = 10\,cm/s, \alpha = -0.02\,fs^{-2}$ and (d) $v_r = 100\,cm/s, \alpha = -0.02\,fs^{-2}$

We have chosen appropriate units for observing the trajectory of the atoms on picosecond time scale. It could be observed that the atoms are getting focused or defocused. It could be deduced from Fig. 8 that the velocity of atoms under the influence of optical dipole force is nearly 6.5 m/s, which is sufficiently consistent with the recent experimental work on the acceleration of neutral atoms [8]. For the experimental realization of the proposed scheme, one may use a train of such pulsed laser fields for efficient focusing and defocusing. It could be observed from Fig. 8(a) and 8 (b) that due to the optical dipole force with up-chirped pulses the atoms are getting focused. So it appears that the optical dipole force is acting like an ultrafast optical lens for the diverging atomic beam. Hence the up-chirped few-cycle laser pulses may be used for the creation of state-selected and focused atomic beam simultaneously. On the other hand, atoms are getting defocused due the defocusing optical dipole force with down-chirped pulses. This could be seen from Fig. 8 (c) and 8(d).

## IV. CONCLUSIONS

To conclude, we have demonstrated the almost complete and highly robust coherent population transfer in sodium atoms with chirped few-cycle-pulse laser fields. It is shown that the direction of the optical dipole force is sensitive to nature of chirping rate but insensitive towards the time dependent detuning from the atomic resonances. It is shown that at larger Rabi frequencies and chirp rates, the optical dipole force becomes highly oscillatory and may not be relevant for practical applications. Our study on the trajectory of atoms subjected to the optical dipole force shows that sodium atoms, co-propagating or even at rest, with the chirped few-cycle-pulse laser fields may be focused or defocused. The proposed scheme may have potential applications in the preparation of atoms in selected quantum state and creation of optical lens for atomic beam. It may also open new perspectives in the focusing and de-focusing of atoms and molecules.

**Acknowledgments**
Authors would like to express their sincere gratitude and thanks to the unanimous reviewers for their positive and critical comments which helped in improving the manuscript to a considerable extent.
P. Kumar would like to thank MHRD, Government of India for a research fellowship.